\documentclass[12pt]{article}
\usepackage{color}
\usepackage[dvips]{graphicx}
\usepackage{dcolumn}
\usepackage{amsmath}
\usepackage{url}

\begin{document}
\Large
\begin{center}
  A New Freely-Downloadable Hands-on Density-Functional Theory Workbook 
  Using a Freely-Downloadable Version of {\sc deMon2k}
\end{center}
\normalsize

\vspace{0.5cm}

\noindent
Nabila B.\ Oozeer\\
{\em Laboratoire de Spectrom\'etrie, Interactions et Chimie th\'eorique
(SITh),
D\'epartement de Chimie Mol\'eculaire (DCM, UMR CNRS/UGA 5250),
Institut de Chimie Mol\'eculaire de Grenoble (ICMG, FR2607),
Universit\'e Grenoble Alpes (UGA)
301 rue de la Chimie, BP 53, F-38041 Grenoble Cedex 9, France} 

\vspace{0.5cm}

\noindent
Abraham Ponra\\
{\em University of Maroua, Cameroon\\
e-mail: abraponra@yahoo.com}

\vspace{0.5cm}

\noindent
Anne Justine Etindele\\
{\em Higher Teachers Training College, University of Yaounde I,
P.O.\ Box 47, Yaounde, Cameroon\\
e-mail: annetindele@yahoo.fr}

\vspace{0.5cm}

\noindent
Mark E.\ Casida\\
{\em Laboratoire de Spectrom\'etrie, Interactions et Chimie th\'eorique
(SITh),
D\'epartement de Chimie Mol\'eculaire (DCM, UMR CNRS/UGA 5250),
Institut de Chimie Mol\'eculaire de Grenoble (ICMG, FR2607),
Universit\'e Grenoble Alpes (UGA)
301 rue de la Chimie, BP 53, F-38041 Grenoble Cedex 9, FRANCE\\
e-mail: mark.casida@univ-grenoble-alpes.fr}

\vspace{0.5cm}

\begin{center}
{\bf Abstract}
\end{center}

One of us (MEC) developed a hands-on workbook for density-functional
theory (DFT) during the summer of 2020.  The idea was to have something
that could be used to provide practical teaching for students at the
Masters or advanced undergraduate level that would be free, could be
used on a student's own personal computer, and would complement formal
course work.  The workbook is also very much intended to encourage
students to explore program options, discover theory limitations,
puzzle out what to do when the program does not work as expected, and
to help students transition to thinking and using quantum chemistry
programs as a researcher might do.  After describing the structure 
of the workbook, we describe how the workbook has been used thus far 
as a teaching tool and as a useful step towards research-level problems.

\section{Introduction}
\label{sec:intro}

\begin{quote}
\noindent
``I hear and I forget.\\
I see and I remember.\\
I do and I learn.''\\
--- A popular proverb frequently attributed to Confuscious
\end{quote}

``Quantum chemistry'' generally refers to the fine art of calculating
the various properties of molecules by solving the Schr\"odinger equation.
It normally involves running programs of various degrees of complexity.
Within quantum chemistry, terms like ``theoretical chemistry'' and 
``computational chemistry'' nowadays sometimes seem to be used interchangeably.
However there is at least a historical difference between the theory behind 
the programs that quantum chemists use and a computational chemistry course
intended to teach users (whose main interest may be experimental synthetic
or physical chemistry) how to do routine calculations with routine programs.
This difference is also reflected in different textbooks which are used to
teach these subdisciplines.  For example, a typical trajectory for an 
advanced physical chemistry student interested in deepening their knowledge
of quantum chemistry might be to learn some basic quantum mechanics from 
a good third year undergraduate physical chemistry book such as McQuarrie and
Simon \cite{MS97}, then use Levine's excellent text \cite{L14} in a fourth-year 
course, followed by Szabo and Ostland \cite{SO96} in a first graduate school
course, possibly followed by Helgaker, J{\o}rgensen, and Olsen \cite{HJO00} 
in an advanced theory course.  Computational
chemistry could be introduced beginning as early as the students third
year of undergraduate studies and might use Cramer's book \cite{C04} or
Jensen's book \cite{J07}, to name just two possibilities.  However 
something seems to be missing that is often supplied only partly by tutorials 
for specific programs.  Specifically, the problem with the tutorials is that 
they typically show how to do calculations without actually having 
students {\em explore program options}, discovering {\em theory limitations}, 
and {\em puzzling out what to do when the program does not work as expected}.  
Students need this to {\em transition to thinking and using quantum chemistry 
programs as a researcher might do}.  This is why one of us (MEC) spent time 
\marginpar{\color{blue} MEC}
in the summer of 2020 developing a freely downloadable workbook 
\cite{workbook1} which will just be referred to here as the Workbook 
(with a capital W).  It uses a freely downloadable {\sc Linux} serial 
\marginpar{\color{blue} {\sc deMon2k}}
version of the {\sc deMon2k} program \cite{deMon2k}.  This means that the
student can download a fully functional quantum chemistry program onto their
personal computer for free and run it there to gain valuable experience
about how such a program works.  This is not intended to replace existent
courses and basic textbooks, nor do we feel that the applications in the
\marginpar{\color{blue} Workbook}
Workbook are particularly specific to {\sc deMon2k}.  (We are not trying
to claim that our software is better than other software for present
purposes.)  Rather we propose this Workbook as a valuable (but freely
available) complement to existing resources.  The goal of the present 
article is to describe our pedagogical experiences with this Workbook 
thus far and where we hope to go in the future.  As we will frequently 
need to refer to specific authors, we will do so by their initials only.
For example, MEC is one of the authors of {\sc deMon2k} and so it was
very natural for the him to base the Workbook on this program.

This article is organized as follows.  The next section says a few words
about {\sc deMon2k} and about the organization of the Workbook.
The Workbook has been $\beta$ tested by NBO and AP, two students at
very different levels.
Section~\ref{sec:nabila} focuses on NBO, an advanced (i.e., 
\marginpar{\color{blue} NBO\\ AP}
third-year undergraduate) $\beta$-tester of the Workbook. 
It does not seem necessary to document the experience of AP who had
fewer obstacles to surmount in using the Workbook.
However, Sec.~\ref{sec:gateway} reports how the Workbook unexpectedly opened
up an avenue for research in Cameroon in Africa, 
where computational resources are significantly more limited than in France,  
\marginpar{\color{blue} AJE}
specifically in the context of the PhD thesis work proposed by AJE for
her thesis student AP.
Section~\ref{sec:conclude} sums up our principle conclusions and hopes
for the future.  

\section{Workbook 1}
\label{sec:workbook}

The Workbook was designed with first-year Masters students in mind who
have already had some introduction to quantum chemistry, perhaps following
the cursus already mentioned in the introduction.  The Workbook is intended
to be complementary to this and to further instruction in quantum chemistry
\marginpar{\color{blue} DFT}
and density-functional theory (DFT) in particular.  Those who are seeking
more information about DFT may be referred to a number of useful basic
texts \cite{PY89,DG90,KH00}.  The reader may also be interested in
a recent perspective article discussing the place of DFT in 
present-day quantum chemistry \cite{THS+22}.
The goal here is not to review either basic quantum
chemistry or DFT more than necessary.  Rather we just seek to provide an 
overview of the Workbook.  This is most conveniently done by
structuring the subsections of this section in the same way 
as the structure of the Workbook.

\subsection{Preliminaries}
The first part of this chapter is just a 

\paragraph{Preface} explaining the objectives of the 
Workbook.  One of these objectives was to help train students new to 
DFT in quantum chemistry, such as NBO and AP.  This is then 
followed by an important part called

\paragraph{Installation} which describes how to download and install 
a serial version of the {\sc deMon2k} program \cite{GCC+12,deMon2k}  
from the {\sc deMon2k} website.  This version should run on any {\sc Linux} 
machine, including personal computers, making it highly convenient for 
self-study and teaching purposes.  Of course, many people do not 
\marginpar{\color{blue} {\sc Linux}}
have {\sc Linux} installed on their personal computer, but this can be 
\marginpar{\color{blue} {\sc VirtualBox}}
circumvented by installing first {\sc VirtualBox} and then the 
\marginpar{\color{blue} {\sc Ubuntu}}
{\sc Ubuntu} version of {\sc Linux} as NBO explains in an appendix 
to the Workbook.  A final part of this chapter is 

\paragraph{Lesson 0: Running the Program}  This involves setting up a 
simple C shell program {\tt run.csh} which takes care of the administration 
of the various files created and used by {\sc deMon2k} by using a simple 
command such as {\tt run.csh O2} which takes information from the file 
{\tt O2.inp} and runs {\sc deMon2k}.  The program {\tt run.csh} is 
deliberately kept simple in order to encourage the student to understand 
and modify it as appropriate.

\subsection{Hydrogen Atom Calculations: Basis Sets and Functionals}  
The first part is 

\paragraph{Lesson 1: The Orbital Basis Set} which
is aimed at familiarizing the student with the concept of the linear 
combination of atomic orbitals (LCAO) approximation,
\marginpar{\color{blue} LCAO\\ MO\\ AO}
\begin{equation}
  \psi_i^\sigma(\vec{r}) = \sum_{\mu} \chi_\mu(\vec{r}) C_{\mu,i}^\sigma \, .
  \label{eq:workbook.1}
\end{equation}
Here $\psi_i^\sigma$ is the $i$th molecular orbital (MO) of spin $\sigma$,
$\chi_\mu$ is an ``atomic orbital'' (AO), and ${\bf C}^\sigma$ is the 
corresponding matrix of MO coefficients.  In practice, the AOs are no longer
solutions of an atomic Schr\"odinger equation but rather are atom-centered
functions chosen for computational convenience as well as for their ability
to cover the variational space.  The most common choice in quantum
chemical programs is to use
gaussian-type orbitals (GTOs) as most of the integrals may be
\marginpar{\color{blue} GTO}
evaluated analytically.  We note in passing that {\sc deMon2k} (or, more
exactly, {\sc deMon} for {\em densit\'e de Montr\'eal}, before it was
renamed) was one of the first programs to use GTOs for DFT and hence to be able
to take advantage of the large amount of quantum chemistry technology
available at that time for calcualtions using GTO basis sets.  Optimization
of the MO coefficients uses the variational principle. Hence this is not
only a lesson about GTO basis sets, but about the proper use of the variational
principle when comparing results from different basis sets. 

This is followed by 

\paragraph{Lesson 2: Density Functionals} which introduces
the numerous density-functional approximations (DFAs) which are available
\marginpar{\color{blue} DFA}
in {\sc deMon2k} (and in many other quantum chemistry programs as well).
Some exact conditions are discussed, notably the self-interaction error,
and an exercise is proposed to test the quality of the different DFAs
for the total and orbital energies of the hydrogen atom where the exact
answer is known.

\subsection{H$_2^+$ and H$_2$: Functionals, Potential Energy Curves, and 
Geometry Optimizations}
The student finally gets to molecules, albeit the simplest possible
molecules.  This chapter of the Workbook begins with 

\paragraph{Lesson 3: The Radical Dissociation Problem} in
which the notorious dissociation problem of H$_2^+$ is discussed in
terms of the particle-number derivative discontinuity present in the
exact exchange-correlation functional, but absent in nearly all DFAs
(with Hartree-Fock an exception, to the extent that Hartree-Fock
may be condered to be a DFA).  As is the case with all of the lessons,
exercises are proposed to help the student obtain a practical understanding
of the non-trivial concept that has been introduced.  Answers to all
the lessons are given in the last chapter of the Workbook.  The student is
likely to encounter their first real difficulty in this lesson --- namely 
the need to modify the sample input file when using meta generalized
gradient approximations.  The needed modification is explained in the
answers in the final chapter and a lengthy discussion is given analyzing and
discussing the results.

The next lesson is 

\paragraph{Lesson 4: Treating Multideterminantal Problems by 
Symmetry-Breaking}.  This is an application to H$_2$ where DFAs work 
reasonably well, provided the trick of symmetry breaking is used.
It is also an excuse to discuss spin coupling and dissociation limits.
Symmetry breaking and spin projection is introduced.  The exercise is 
more challenging because DFT calculations do not necessarily automatically
converge to symmetry-broken solutions unless the user is creative in 
how the program is run.  The problem of self-consistent field (SCF)
\marginpar{\color{blue} SCF}
convergence is also typically particularly severe near symmetry-breaking
points.  This is a good chance for students to explore the different
tricks for making their calculations converge.

The final lesson in this chapter is 

\paragraph{Lesson 5: Analytic Gradients and
Geometry Optimization}.  Variational calculus is introduced as a formal
tool and then used to derive the Hellmann-Feynman theorem which fails 
for atom-centered GTO basis sets.  It is then explained why Pulay forces
need to be introduced and why, in practice, these require good SCF
convergence.  A simple potential energy surface walking algorithm
is introduced and discussed.  The exercises show that the walking algorithms
used in {\sc deMon2k} are actually much better than the simple algorithm
that was proposed.

\subsection{Lesson 6: Singlet Oxygen, $^1$O$_2$}

This chapter is a bit unusual because it was intended as a lesson 
on the multiplet sum method to help AP get started with his PhD 
thesis project involving reactive oxygen species.
It is a nice lesson in group theory applied to multideterminantal wave 
functions.  This lesson would grow into a research paper \cite{PEMC21}
comparing how well
different functionals describe the ground and lowest excited states of 
diatomic oxygen as further described in Sec.~\ref{sec:gateway}.

\subsection{Answers}

This contains answers to the exercises proposed in the various lessons
as well as in-depth analysis and discussion.  It is hoped that the student
will only look at this {\em after} trying the exercises on their own.
It is possible that the student will find better solutions to difficulties
encountered than the ones given in this chapter.

\subsection{Installing {\sc Linux} on a Mac Notebook}

This appendix, written by NBO, describes how she was able to install {\sc
Linux} on her Mac Notebook by first installing {\sc VirtualBox} and then 
installing {\sc Ubuntu}.

\section{Advanced Undergraduate Experience}
\label{sec:nabila}

Early in 2022, NBO volunteered to $\beta$ test the Workbook as part of
a one-month research internship as a third-year undergraduate in
chemical physics.  Her experience is summarized in her report (in
French) which may be found on-line \cite{Nabila}.  In the jargon
of computer programming, $\alpha$ testing refers to debugging (and
other tests) performed by the programmers, while $\beta$ testing refers
to additional testing made by users which may turn up new errors
or just show where further program development is useful to meet their
needs.  Similarly we can say that both AP and NBO were $\beta$ testers
for the Workbook.  However NBO was testing it as someone very new to
to quantum chemistry --- namely as an undergraduate who was just taking
her first quantum physics course at the same time as her one-month
research internship.  We were very curious to see if she could benefit
from the Workbook.  Given that her background at that time was about
the minimum needed, we focused on Lessons 0 and 1.  

Lesson 0 was to see if she could install
and run {\sc deMon2k} on her Mac notebook.  She passed this first hurdle 
with flying colors and contributed a key appendix on the subject to the 
Workbook as described in Sec.~\ref{sec:workbook}.

Here we focus on her work with Lesson 1.  This Lesson was
considerably simplified by focusing only on the problem of constructing 
GTO basis sets for solving the hydrogen atom as exactly as possible and
comparing against the known exact solutions.  In order to be able to treat
this one-electron system as exactly as possible, NBO carried out 
self-interaction error free Hartree-Fock calculations using the {\tt FOCK} 
option. NBO was then free to focus on the details of GTO basis sets 
and on the variational principle.  For reasons of simplicity, Hartree
atomic units ($\hbar = e = m_e = 1$) will be used throughout this section.

She began with the classic textbook exercise of finding the variationally 
optimal exponent of a single GTO,
\begin{equation}
  \psi_{\mbox{trial}}(\vec{r}) = \left( \frac{2\alpha}{\pi} \right)^{3/4}
  e^{-\alpha r^2} \, .
  \label{eq:nabila.1}
\end{equation}
The radial hamiltonian is,
\begin{equation}
  \hat{h} = -\frac{1}{2r} \frac{\partial^2}{\partial r^2} r
  - \frac{1}{r} + \frac{1}{2r} \hat{L}^2 \, ,
  \label{eq:nabila.2}
\end{equation}
where $\hat{L}$ is the angular momentum operator which acts as an
annihilation operator on the $s$-type function $\psi_{\mbox{trial}}$.
From this,
it is then a straightforward exercise (involving gaussian integrals
which, by coincidence, NBO was also studying in her quantum physics class
at about the same time) to determine that the variational integral is 
\begin{equation}
  W(\alpha) = \frac{3}{2} \alpha - 2 \sqrt{\frac{2\alpha}{\pi}} \, .
  \label{eq:nabila.3}
\end{equation}
Taking derivitives and setting equal to zero leads to analytic expressions
for the variationally optimal value of $\alpha$ and the associated
energy:
\begin{eqnarray}
  \alpha_{\mbox{min}} & = & \left( \frac{2}{3} \right)^2 \mbox{2}{\pi}
   = 0.2829 \nonumber \\
  E_{\mbox{min}} & = & W(\alpha_{\mbox{min}}) = -0.4244 \, .
  \label{eq:nabila.4}
\end{eqnarray}
\begin{figure}
        \begin{center}
\includegraphics[width=0.8\textwidth]{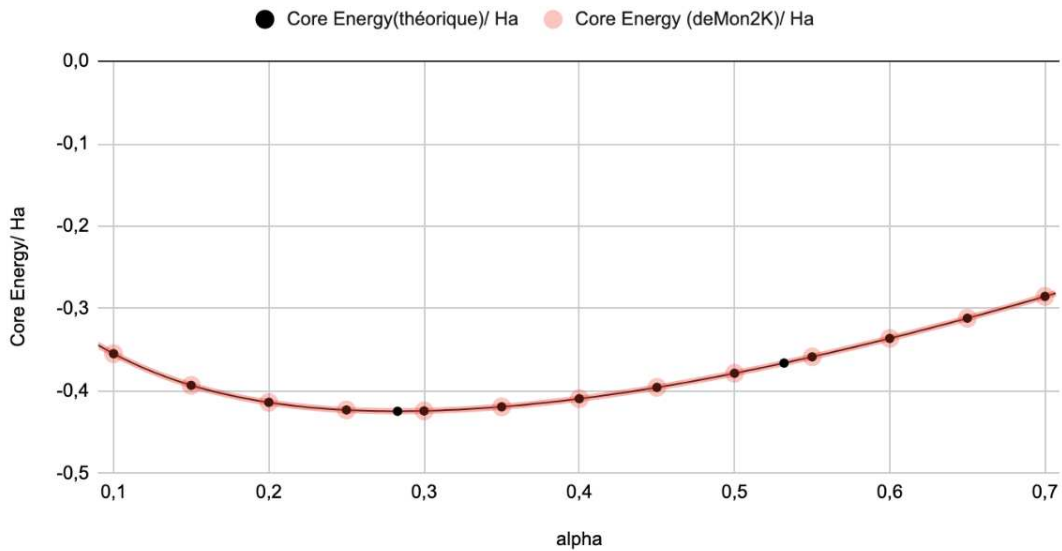}
        \end{center}
\caption{
Graph of the variational integral [Eq.~(\ref{eq:nabila.3})] as 
a function of the exponent $\alpha$ as calculated by {\sc deMon2k}.
Black dots: analytic values derived by NBO.  Peach colored dots: calculated
by {\sc deMon2k}.  We note that ``Core Energy'' here means the total energy
neglecting electron repulsions.
\label{fig:1G}
}
\end{figure}
It was very important for NBO to be able to verify this explicitly by
creating her own single function GTO in the {\sc deMon2k} {\tt BASIS}
file where she could modify the exponent $\alpha$.  This resulted in 
the graph shown in {\bf Fig.~\ref{fig:1G}} which comforted NBO by showing
that the program gave the same results as she had found.  This, of course,
occurs because of the use of explicit formulae for gaussian integrals 
within the {\sc deMon2k} program.

\begin{figure}
        \begin{center}
\begin{tabular}{cc}
(a) & \\
(b) & \includegraphics[width=0.6\textwidth]{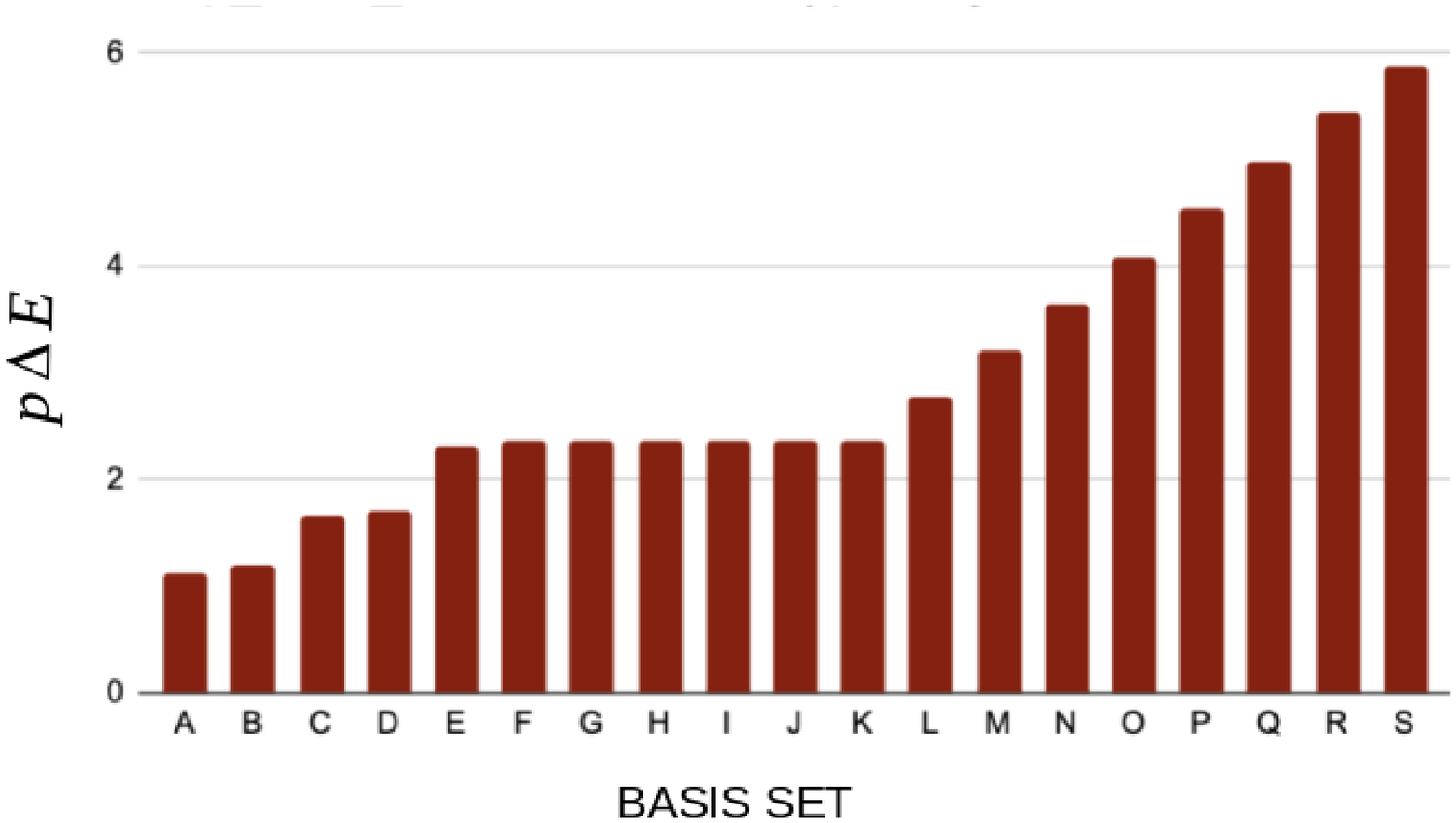} \\
(c) & \includegraphics[width=0.6\textwidth]{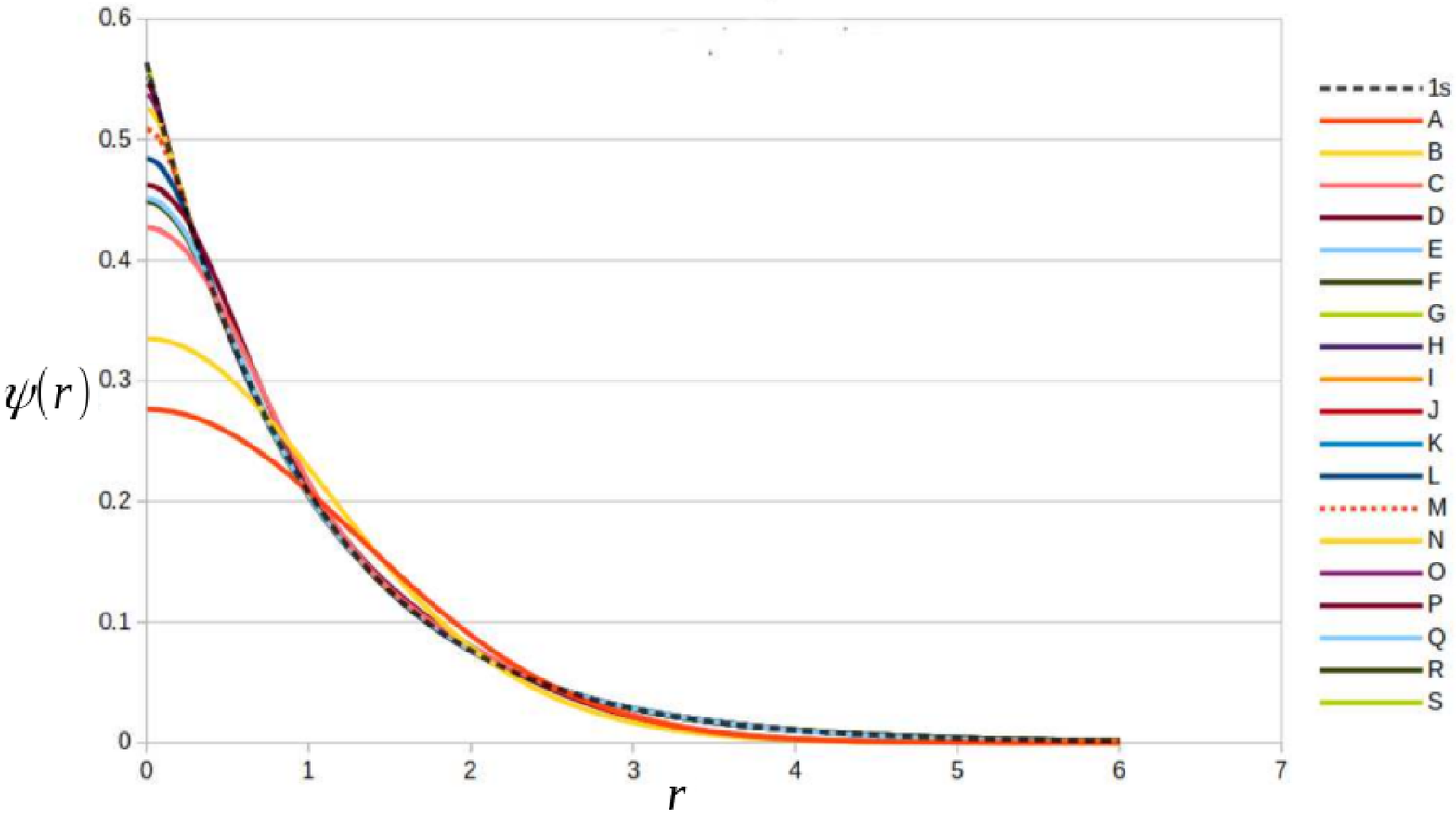} \\
    & \includegraphics[width=0.6\textwidth]{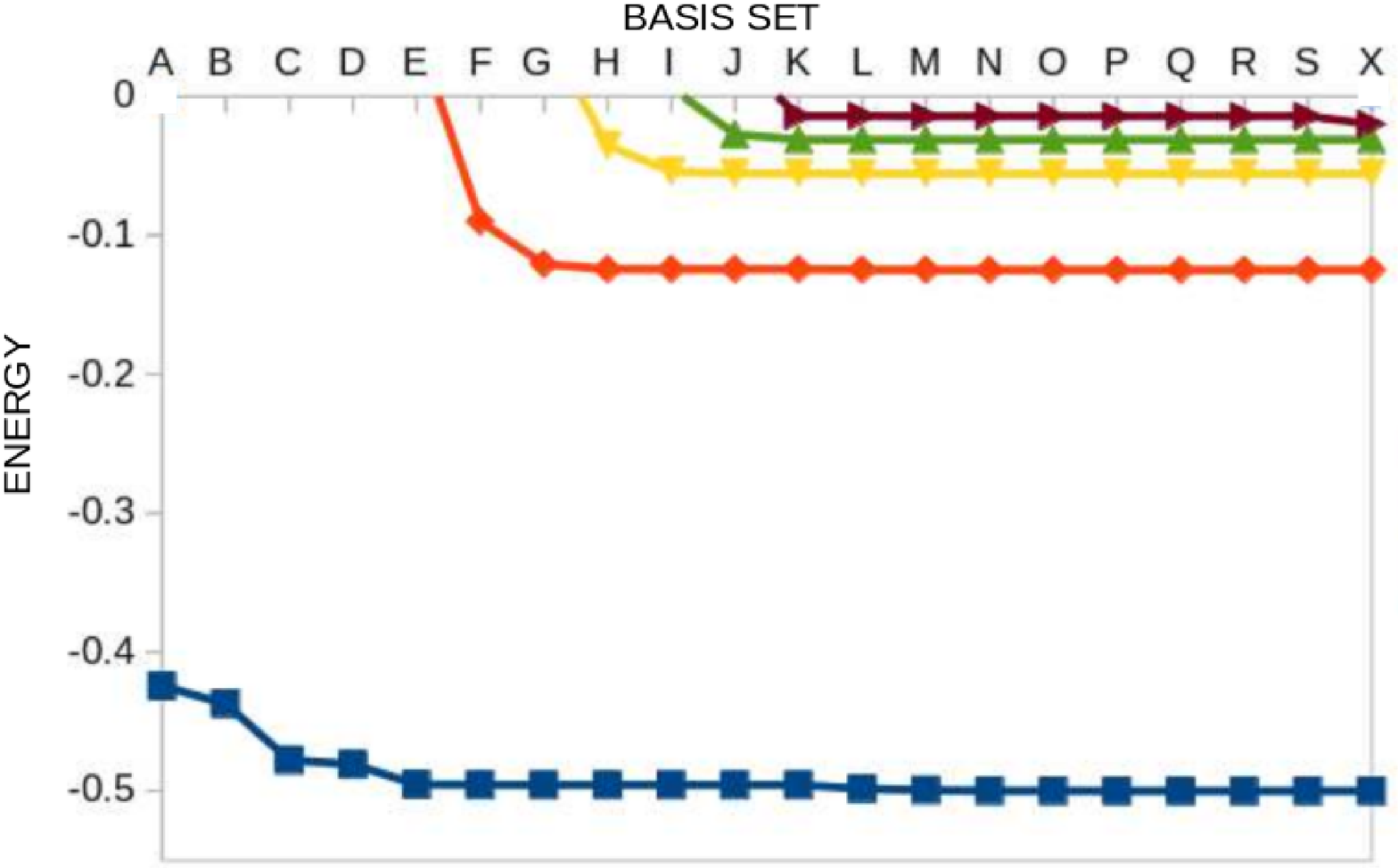} 
\end{tabular}
        \end{center}
\caption{
Results of NBO's numerical experiments with the even-tempered basis
sets that she constructed herself: 
(a) groundstate energy error $p\Delta E = -\log_{10} (E - E_{\mbox{exact}})$ 
for the different basis sets, (b) convergence of the wave function,
and (c) convergence of excited state energies (X stands for ``exact'').
\label{fig:eventempered}
}
\end{figure}

The next step was to go beyond the sort of exercise normally found in textbooks
to something looking more like research.  The ``research question'' was,
\begin{quote}
  \noindent
  How can a GTO basis be constructed that systematically converges to the
  exact answer and how large does it have to be for a given level of accuracy?
\end{quote}
To this end, NBO constructed a series of even-tempered basis sets \cite{FR79,SR79,KW03}, 
meaning that the GTOs that she put in the {\tt BASIS} file were sets of $s$-type GTOs 
of the form of Eq.~(\ref{eq:nabila.1}) but with a series of 
$\alpha_i = \alpha_0 \beta^{i-1}$ for integer $i$. 
Even-tempered basis sets are known to optimally cover variational space
in a certain sense, making them a good choice.
She chose $\alpha_0=0.00221015625$ and $\beta=2$ with $i=1,2,3,\cdots,19$, so that 
$\alpha_8=\alpha_{\mbox{min}}$ of Eq.~(\ref{eq:nabila.4}) obtained by 
her hand calculation plus added tight and diffuse functions.  
Results are shown in {\bf Fig.~\ref{fig:eventempered}}.
Part (a) of the figure shows that the energy descreases significantly when NBO added
more and more tight functions (A-D) and also when she added a first more diffuse
function (E), but not when she continued to add more diffuse functions (F-K).  After that
NBO added more ands more tight functions (J-S) until agreement with the exact energy
was nearly 10$^{-6}$ Ha = 0.0006 kcal/mol.  Part (b) of the figure shows that the wave
function is also converging, with an improved description of the long range and especially
of the cusp behavior.  Part (c) of the figure tests a consequence of what physicists
call the Hylleraas-Undheim-MacDonald theorem and what mathematicians call Cayley's
interleaving theorem which says that adding additional basis functions in a linear
variational problem can only result in a new set of eigenvalues that interleave the
old eigenvalues.  In the present context, this means that every $n$th energy is an
upper bound to the true $n$th energy level.  In contrast to the ground-state situation,
and as expected, we see that adding diffuse functions is essential for proper convergence 
of excited-state energies.

The above description of what NBO did is not an exhaustive list of everything done during
her short, but very productive, one-month internship.  What is more important is that she
was able to do quite a lot, in no small part, because of the availability of the Workbook
as a learning tool and because she could work directly with her own personal computer.  
She even went beyond what was described in the Workbook to make her own exercises.
We believe this is very much in the spirit of the Workbook which aims at providing tools,
and the learning environment needed, for self-learning and moving towards research 
using quantum chemistry.  But let us translate NBO's own conclusions directly from 
her internship report:
\begin{quote}
  \noindent
  ``The advantage of the software used and of the Workbook is that a student can 
  learn and progress by themselves in learning DFT armed with only their computer.
  For this reason, I can recommend the Workbook to other students interested in
  a hands-on approach to basic quantum chemistry.''
\end{quote}

\section{Gateway towards Research}
\label{sec:gateway}

This section describes how the Workbook became a gateway towards research-level
work performed on computers in Cameroon and published in a peer-reviewed
research journal \cite{PEMC21}.  This is significant because of the high-level
of interest in theoretical research in Africa, but a real lack of computational
resources combined with an, at least partially incorrect, perception that such
resources are needed to do cutting edge research.  AJE and MEC are happy to
\marginpar{\color{blue} ASESMA}
acknowledge the African School on Electronic Structure Methods and 
Applications (ASESMA), with which we have been involved since 2012,
for helping to create a research environment in Africa where efforts such
as our own are possible.

We are also happy to learn that Jim Gubernatis will receive the 2023 
John Wheatley Award of the American Physical Society, 
``For ongoing commitment to developing physics in Africa through initiating 
the African School on Electronic Structure Methods and Applications and 
leadership in bringing together African physicists from across the continent 
to create a Pan-African physics communication vehicle.''
ASESMA was established to improve the research environment
for Africans doing theoretical research in Materials Science
and had its first meeting in 2008 \cite{CMS10,F11,AM15,M16,ACH+18}.
quantum chemistry has been a part of ASESMA from very early in the history
of the school.  In fact, theoretical solid-state physics and quantum chemistry 
have enormous overlaps, even if their language and approximations differ.  
This difference has been diminishing over the years as solid-state physics 
focuses more on nanostructures and as quantum chemists turn towards ever 
larger systems.  Partly for this reason, and partly because of the lack of 
anything analogous to ASESMA for quantum chemists in Africa, there has 
always been a subgroup of ASESMA theoreticians interested in quantum chemistry.

ASESMA students and African mentors still have to face substantial problems.
A bottleneck in Africa has been scarce computing resources as many African 
countries, such as Cameroon, provide little governmental support for research.
This is partly made up by access to computing resources via collaborations 
with researchers in developed countries and partly by limited access to high 
performance computing centers in wealthier African countries, notably 
South Africa.  But, we feel that this does not replace the need for local 
computational resources for education and for research which should not 
always require large amounts of computational resources.  
This is one place that the Workbook can be very helpful as it is based upon 
freely downloadable software that will run on any {\sc Linux}
system.  AJE and AP already had some experience with solid-state quantum 
physics codes which typically use very different types of basis sets than 
do quantum chemistry codes.  As the thesis subject proposed by AJE for AP 
focuses on molecules in a biological context, AP needed to be trained in 
different methods.  This was one of the reasons (but not the only one!)
that MEC developed the Workbook in the first place.  It has been used by 
AJE and MEC to help train AP and the Workbook is actujally subtitled 
``Abraham's Workbook'' as he was the first $\beta$ tester.  

Rather than describe AP's $\beta$ testing experience (which was positive),
we focus instead on how the Workbook became, almost accidently, a gateway
to computational work that could be carried out locally using only the 
limited computational resources available locally in Cameroon and
published in an archival research journal \cite{PEMC21}.  

The research project proposed by AJE as AP's thesis project concerns
the use of buckminsterfullerene (C$_{60}$) and its derivatives as 
antioxidants that may be used to protect the body and the environment 
\marginpar{\color{blue} ROS}
from highly reactive oxygen species (ROSs).  These later include a variety of 
small molecules.  Although ground-state $^3$O$_2$ (which is in a triplet
electronic state) is a notoriously benign biradical, for kinetic reasons, 
it can easily form the the highly-reactive superoxide $^\bullet$O$_2^-$ by 
reduction or the equally highly-reactive (biradical) singlet oxygen $^1$O$_2$ 
by exposure to sunlight in the presence of a suitable photosynthetizer.
Other ROSs include, but are not limited to, hydrogen peroxide H$_2$O$_2$
and the hydroxyl radical $^\bullet$OH.  ROSs play a role in 
atmospheric chemistry, water purification, and occur naturally in biology
during photosynthesis and during respiration in mitochondria.  Needless to
say, living things have mechanisms to keep ROSs from getting out of hand, but
sometimes extra help is needed.  Buckminsterfullerene is so reactive towards
radicals that it has been called a ``radical sponge'' \cite{MML92} and has
even been commercialized under this name for use in cosmetics \cite{Sponge}.
The idea then, was to investigate theoretically the reactivity of C$_{60}$
with ROSs.  This led naturally to a discussion of the low-lying excited
states of O$_2$ and inspired Lesson 6 of the Workbook which explains how
group theory and the multiplet sum method can be applied to determine
the ground- and excited-state energies of O$_2$.  Thus Lesson 6 was designed
in the first instance to aid a student to understand a particular approach
to a problem.

What happened next was a combination of things, one of which was the 
very unexpected realization that much of the information on the internet 
and even in many research articles gives an incorrect, or at least misleading, 
description of the ground- and electronic-excited states of O$_2$.
It was as if the lessons of Herzberg had been forgotten only a few
generations latter!  This can explain, at least in part, why only a very 
few DFT calculations have been done which properly treat the symmetry of 
O$_2$ electronic states.  It was clearly time to remind the world again
of a little group theory (that was in Lesson 6) and to test out which are  
the best functionals for treating the O$_2$ multiplet problem (a research-level
problem do-able with limited computer resources).  This is not the place to 
repeat this work, but rather to just review what seem to be the important
points.

\begin{figure}
        \begin{center}
\includegraphics[width=0.8\textwidth]{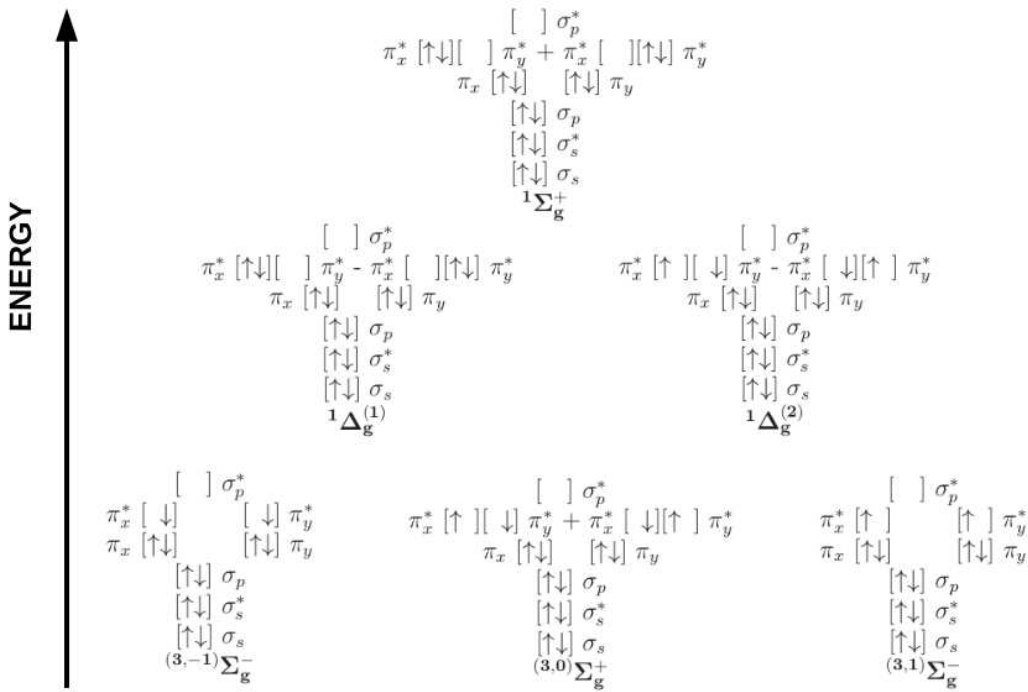}
        \end{center}
\caption{
MO diagrams and multiconfigurational nature of the
ground triplet state and two lowest singlet excited states of
O$_2$.  These schematic diagrams are taken directly from the Workbook.
\label{fig:O2}
}
\end{figure}
First of all, the Workbook lessons are presented as a combination of teaching
and of self-exploration.  For those of us who have been involved in ASESMA,
this is an echo of the organization where the first week is devoted to learning
concepts and tools while the second week is devoted to a miniproject which
might or might not lead to a publication.  Lesson 6 was not intended to lead
to a publication but the exercises do have the form of a miniproject.
{\bf Figure~\ref{fig:O2}} shows MO diagrams for the ground- and two lowest
excited-states of O$_2$ using {\em real} orbitals.  The symmetry-imposed
configuration mixing is also shown in a schematic fashion.  This and a program
such as {\sc deMon2k} which allows MO occupation numbers to be fixed 
are enough to be able to calculate the energies of all the 3 states from 
single-determinantal states made from some reference state which we took 
to be a state having an equal fractional occupation number of half 
of each spin in each of the two $\pi_x^*$ and $\pi_y^*$ orbitals.  
Most quantum chemistry programs have some option which allows the 
creation of this type of reference state, but not every quantum chemistry 
program allows the changing of MO occupation numbers.  It is then a 
``simple matter'' to calculate the three state energies,
\begin{eqnarray}
  E[X\, ^3\Sigma_g^-] & = & E(\pi_x^*[\uparrow, \,\,\,\,][\uparrow, \,\,\,\,]\pi_y^*) \nonumber \\
  E[a\, ^1\Delta_g] & = & E(\pi_x^*[\uparrow, \,\,\,\,][\uparrow, \,\,\,\,]\pi_y^*) + 2P \nonumber \\
  E[b\, ^1\Sigma_g^+] & = &  E(\pi_x^*[\uparrow, \,\,\,\,][\uparrow, \,\,\,\,]\pi_y^*) + 2F \, ,
  \label{eq:gateway.1}
\end{eqnarray}
where
\begin{equation}
  F = E(\pi_x^*[\uparrow, \,\,\,\,][\,\,\,\, , \downarrow]\pi_y^*) - E(\pi_x^*[\uparrow, \,\,\,\,][\uparrow, \,\,\,\,]\pi_y^*)
  \label{eq:gateway.2}
\end{equation}
is the spin-flip energy and 
\begin{equation}
  P = E(\pi_x^*[\uparrow, \uparrow][\,\,\,\, , \,\,\,\,]\pi_y^*) - E(\pi_x^*[\uparrow, \,\,\,\,][\uparrow, \,\,\,\,]\pi_y^*) 
  \label{eq:gateway.3}
\end{equation}
is the spin-pairing energy.
Previous work on octahedral Fe(II) spin-crossover complexes had shown that
the spin-pairing energy ($\tilde{P}=P-F$) can be very sensitive to the 
choice of density-functional approximation (DFA)
\cite{FMC+04,FCL+05,GBF+05,LVH+05,ZBF+07,LC22}, 
suggesting that the difference of the $^1\Sigma_g^-$ and $^1\Delta_g$
energies was going to be very sensitive to choice of DFA, but might
\marginpar{\color{blue} OLYP}
be better approximated by the use of the OLYP \cite{LYP88,HC01} 
generalized gradient approximation (GGA) than by the use of other GGAs.  
\marginpar{\color{blue} GGA}

As no general study of DFAs for $^1$O$_2$ energies had yet been carried out,
this was an excellent opportunity to contribute new knowledge to the chemical
literature \cite{PEMC21}.  Figure 13 of Ref.~\cite{PEMC21} did indeed
confirm that the OLYP error in $\tilde{P}$ was among the lowest of the
various GGAs tried.  More importantly Fig.~12 of Ref.~\cite{PEMC21} shows
that the $^1\Delta_g$ excitation energy is less sensitive to the choice of
DFA than is the $^1\Sigma_g^-$ excitation energy, meaning that spin-flip 
energies are easier to describe in DFT than are spin-pairing energies, at
least in this application.  We were thus able to contribute something to
the knowledge of how different DFAs are able to treat spin.  In the end,
we found that we had to go fairly high up the Jacob's ladder \cite{PS01,PRC+09} 
of DFAs to at least meta-GGAs or even hybrid meta-GGAs in order to get 
the best level of accuracy \cite{PEMC21}.  Although our paper was only 
published a year ago, it has already received several citations,
meaning that we can make an impact with the Workbook, a freely downloadable
serial version of {\sc deMon2k}, and the level of computational resources
available locally in Cameroon.

\section{Conclusion}
\label{sec:conclude}

A freely-downloadable hands-on density-functional theory Workbook
that works with a\\ freely-downloadable serial version of {\sc deMon2k},
executable on any {\sc Linux} operating system, has been described.
This is a learning tool for serious researchers at the level of beginning
Masters students in chemistry or beyond, though we have described the
successful use of the Workbook by a third year undergraduate student.
It is particularly useful for those who wish to learn at home or
who have limited access to computing resources.  A specific objective
is to create an opening towards research in quantum chemistry
precisely by {\em not} avoiding practical and formal difficulties, but
rather by encouraging and helping the student to come to terms with these
difficulties.

We think that this Workbook may be especially valuable in the developing
world and give an example from Cameroon.  Governments in developing 
countries frequently have little or no funds for scientific research
except possibly for the most basic sort of applied research such as
water purity analysis.  However researchers in these countries are often
hungry to do more basic research, but may feel discouraged because of their
limited resources.  This Workbook is also a wake-up call to those researchers
that it is possible to advance science with only limited resources 
{\em provided} (and this is the key point) they have enough knowledge to
know what are the important fundamental problems and how to contribute
to solving them.  We would like to think that the Workbook is part of 
what is needed to enable researchers in both developing and developed
countries to educate themselves and find where they can contribute to 
the advancement of science.  We have given one example, but we hope that
there will be many others in the coming years.

\section*{Acknowledgements}

Prof.\ Ponnadurai Ramasai is gratefully acknowledged for encouraging
us to write up and submit this manuscript.
{\sc deMon2k} is a group effort and MEC is
grateful to the other deMon developers without whom none of this
work would have been possible.
NBO thanks Pierre Girard for valuable help with the installation of
{\sc VirtualBox} and {\sc Linux}. MEC and AJE wish to acknowledge
many useful interactions through ASESMA which have helped to make
this work possible. MEC would also like to acknowledge the organizers
of Virtual Winter School on Computational Chemistry (VWSCC) whose
workshop ``Africa Calling'' drew our attention to the Virtual Conference 
on Chemistry and its Applications (VCCA-2022).

\section*{Author Contributions}

Author credit has been assigned using the CRediT contributor roles taxonomy system \cite{CRediT}.
\\
{\bf Nabila B. Oozeer}: Investigation, writing - review \& editing, visualisation.
\\
{\bf Abraham Ponra}: Investigation, writing - review \& editing, visualisation.
\\
{\bf Anne Justine Etindele}: Supervision, writing - review \& editing.
\\
{\bf Mark E.\ Casida}: Conceptualization, project administration, supervision, writing - orginal draft, formal analysis, visualisation.

\newpage

\end{document}